\documentclass[prd,twocolumn,amsmath,amssymb,floatfix,superscriptaddress,nofootinbib]{revtex4-2}

\usepackage{graphicx}
\usepackage[colorlinks=true, linkcolor=blue, citecolor=blue, urlcolor=blue]{hyperref}
\usepackage{amsmath}
\usepackage{amssymb}
\usepackage{graphicx}

\begin{document}
	
\title{Approximate N$^5$LO Higgs boson decay width $\Gamma(H\to\gamma\gamma)$}

\author{Yu-Feng Luo}
\email{luoyf@stu.cqu.edu.cn}
\author{Jiang Yan}
\email{yjiang@cqu.edu.cn}
\author{Zhi-Fei Wu}
\email{wuzf@cqu.edu.cn}
\author{Xing-Gang Wu}
\email{wuxg@cqu.edu.cn}

\affiliation{Department of Physics, Chongqing Key Laboratory for Strongly Coupled Physics, Chongqing University, Chongqing 401331, P.R. China}
	
\date{\today}
	
\begin{abstract}	

The precision and predictive power of perturbative QCD (pQCD) prediction depends on both a precise, convergent fixed-order series and a reliable way of estimating the contributions of unknown higher-order (UHO) terms. It has been shown that by applying the Principal of Maximum Conformality (PMC), which applies the renormalization group equation recursively to set the effective magnitude of $\alpha_s$ of the process, the remaining conformal coefficients will be well matched with the corresponding $\alpha_s$ at each orders, leading to a scheme-and-scale invariant and convergent perturbative series. Thus different from conventional scheme-and-scale dependent fixed-order series, the PMC series will provide a more reliable platform for estimating UHO contributions. In this paper, by using the total decay width $\Gamma(H\to\gamma\gamma)$ which has been calculated up to N$^4$LO QCD corrections, we derive its PMC series by using the PMC single-scale setting approach and estimate its unknown N$^5$LO contributions by using the Bayesian analysis. The Bayesian-based approach estimates the magnitude of the UHO contributions based on an optimized analysis of probability density distribution, and the predicted UHO contribution becomes more accurate when more loop terms have been known to tame the probability density function. Using the top-quark pole mass $M_t$=172.69 GeV and the Higgs mass $M_H$=125.25 GeV as inputs, we obtain $\Gamma(H\to\gamma\gamma) =9.56504~{\rm keV}$ and the estimated N$^5$LO contribution to the total decay width is $\Delta\Gamma_H=\pm1.65\times10^{-4}~{\rm keV}$ for the smallest credible interval of $95.5\%$ degree-of-belief.

\end{abstract}

\maketitle

\section{Introduction}

The ATLAS and CMS collaborations have discovered the Higgs boson in 2012~\cite{ATLAS:2012yve, CMS:2012qbp}, being consistent with the elementary particle suggested by Standard Model (SM). The Higgs boson answers some of the most profound questions in physics, such as how the masses of the elementary particles and the $W^{\pm}/Z^{0}$ gauge bosons come from, how the electroweak phase transition governs the evolution of the early universe, and etc. It is then crucial to verify and study the Higgs properties, either experimentally or theoretically.

Precise measurements of the Higgs boson production and decay channels provide critical tests of the SM and are vital in the exploration of possibly new physics beyond the SM. Over the past decade, since its discovery, many new measurements on the Higgs boson properties have been performed by the collaborations at the LHC. Some new Higgs factories such as the International Linear Collider (ILC)~\cite{ILC:2013jhg}, the Circular Electron Positron Collider (CEPC)~\cite{CEPCStudyGroup:2018ghi} and the Future Circular Collider~\cite{FCC:2018byv} have been designed to further improve the experimental precisions on the Higgs properties. The Higgs boson is being moved from the object of a search to an exploration tool. Till now almost all of the related measurements are in agreement with the SM predictions within errors. As one of the most important decay channels of the Higgs, it has been shown that the process $H\to\gamma\gamma$ has an observable fraction $(2.50\pm0.20)\times 10^{-3}$~\cite{ParticleDataGroup:2022pth}, which plays an important role in Higgs phenomenology.

Because the photon is massless, the process $H\to \gamma\gamma$ is a loop-induced process even at the leading order level, whose amplitude can be decomposed into a bosonic contribution, stemming from the $W$ boson, and the fermionic contributions, respectively. More explicitly, its decay width can be written as
\begin{align}
	\label{Gamma}
	\Gamma(H\rightarrow\gamma\gamma)=\frac{M_H^3}{64\pi} \left |A_W+\sum_{f}A_f \right|^2,
\end{align}
where $M_H$ is the Higgs mass, $A_W$ is the contribution from the purely bosonic diagrams, and $A_f$ is the contribution from the amplitudes with $f=(t,b,c,\tau)$, which corresponds to the top quark, the bottom quark, the charm quark, and the $\tau$ lepton, accordingly. The above equation can be further rewritten as~\cite{Yu:2018hgw}
\begin{align}
	\label{GammaH}
	\Gamma(H\rightarrow\gamma\gamma)=\frac{M_H^3}{64\pi} \left(A_{\rm LO}^2+A_{\rm EW}\frac{\alpha}{\pi}\right) + R_{n},
\end{align}
where $\alpha$ is the fine-structure constant, $A_{\rm EW}$ is the electroweak (EW) correction~\cite{Ellis:1975ap, Shifman:1979eb}, $A_{\rm LO}$ is the leading-order (LO) contribution, and $R_{n}$ represents the QCD corrections in which $n$ represents the QCD correction has been calculated up to $n_{\rm th}$-loop level. At the present, the LO, the next-to-leading order (NLO), the N$^2$LO, the N$^3$LO, and the N$^4$LO perturbative QCD (pQCD) corrections for $\Gamma(H\to\gamma\gamma)$ have been done in Refs.\cite{Ellis:1975ap, Shifman:1979eb, Zheng:1990qa, Dawson:1992cy, Djouadi:1990aj, Djouadi:1993ji, Melnikov:1993tj, Inoue:1994jq, Spira:1995rr, Fleischer:2004vb, Harlander:2005rq, Anastasiou:2012hx, Maierhofer:2012vv, Sturm:2014nva, Marquard:2016dcn, Actis:2008ts}. Especially, the fermionic contribution which forms a gauge-invariant subset has been calculated up to N$^4$LO level~\cite{Sturm:2014nva}. Those improvements give us good basis to achieve precise pQCD prediction on $\Gamma(H\to\gamma\gamma)$. On the other hand, future precise measurements on the Higgs boson decays may determine the branching fraction of its decay into two photons up to a high precision of one percent level~\cite{FCC:2018vvp}. Thus to fully exploit future precise measurements, it is important to achieve high precision theoretical prediction as much as possible, as is the purpose of the present paper.

\section{The four-loop prediction $R_4$ under the PMC and the higher-order contribution using the Bayesian analysis}

The perturbative series of the QCD correction $R_{4}$ up to ${\cal O}(\alpha^5_s)$-level can be read from Refs.\cite{Maierhofer:2012vv, Sturm:2014nva}, which is given in $n_f$-series with $n_f$ being the active number of quark flavors. For later convenience of applying the renormalization group equation (RGE) to set the effective magnitude of $\alpha_s$, we reexpress it as a $\{\beta_i\}$-series by using the general degeneracy relations of the QCD theory among different orders~\cite{Bi:2015wea}, e.g.
 \begin{align}	
 R_{4}=&\sum_{i=1}^{4} r_i(\mu_r^2/Q^2) a^{i}(\mu_r)  \label{rhoConv1} \\
 =&r_{1,0}a(\mu_r)+[r_{2,0}+\beta_0 r_{2,1}] a^2(\mu_r) \notag \\
  &+[r_{3,0}+\beta_1 r_{2,1}+2\beta_0 r_{3,1}+\beta_0^2 r_{3,2}] a^3(\mu_r) \notag \\
  &+[r_{4,0}+\beta_2 r_{2,1}+2\beta_1 r_{3,1}+\frac{5}{2}\beta_1 \beta_0 r_{3,2} + 3\beta_0 r_{4,1} \notag \\
  &+3\beta_0^2 r_{4,2}+\beta_0^3 r_{4,3}] a^4(\mu_r)+ {\cal O}(a^5),\label{rhoConv2}
\end{align}
where $a=\alpha_s/\pi$ and $Q=M_t$ ($M_t$ being the top-quark pole mass), which represents the typical momentum flow of the process.  The $\{\beta_i\}$-functions have been calculated up to five-loop level in the $\overline{\rm MS}$-scheme~\cite{Baikov:2016tgj}. The expansion coefficients $r_{i,j}$ in Eq.(\ref{rhoConv1}) can be derived from the ones of Refs.\cite{Maierhofer:2012vv, Sturm:2014nva} via proper transformations. In Refs.\cite{Maierhofer:2012vv, Sturm:2014nva}, the perturbative expressions are given in the form of the $\overline{\rm MS}$-scheme top-quark running mass $(m_t)$. Following the arguments of Ref~\cite{Wang:2013bla}, we will transform it into the perturbative series over the top-quark pole mass ($M_t$) with the help of the  $O(\alpha_s^5)$-level relation between $m_t$ and $M_t$~\cite{Marquard:2016dcn} in order to avoid the confusion of applying the PMC scale-setting procedures, e.g. only the RGE-involved ${\beta_i}$-terms are remained and adopted for fixing the correct magnitude of the strong coupling and its argument, e.g. the PMC scale $Q^*$. The coefficients $r_{i,0}$ are conformal ones which are free of renormalization scale $\mu_r$, and the non-conformal coefficients $r_{i,j(\neq0)}$ are functions of $\mu_r$ which can be reexpressed as
\begin{align}
	r_{i,j}=\sum_{k=0}^{j}C^k_j\hat{r}_{i-k,j-k}\ln^k(\mu_r^2/Q^2),
\end{align}
where $\hat{r}_{i,j}=r_{i,j}|_{\mu_r=Q}$. The RGE determines the running behavior of $\alpha_s$ and is scheme dependent. By applying the Principal of Maximum Conformality (PMC)~\cite{Brodsky:2011ta, Brodsky:2011ig, Brodsky:2012sz, Brodsky:2012rj, Mojaza:2012mf, Brodsky:2013vpa}, which applies the RGE recursively to set the effective magnitude of $\alpha_s$ of the process, the remaining conformal coefficients will be well matched with the corresponding $\alpha_s$ at each orders, leading to a scheme-and-scale invariant and convergent perturbative series free of divergent renormalon terms~\cite{Wu:2013ei, Wu:2014iba, Wu:2019mky, DiGiustino:2023jiq}. The PMC reduces in the Abelian limit to the Gell-Mann-Low method~\cite{Gell-Mann:1954yli} and it provides a solid way to extend the well-known Brodsky-Lepage-Mackenzie (BLM) method~\cite{Brodsky:1982gc} to all orders.

The PMC single-scale approach (PMCs)~\cite{Shen:2017pdu, Yan:2022foz} determines an overall effective $\alpha_s$ (its argument is called as the PMC scale) for the fixed-order predictions, and the resultant perturbative series provide a good basis for demonstrating that the PMC series is free of renormalization scale-and-scheme ambiguities up to any fixed order~\cite{Wu:2018cmb}, being consistent with the fundamental renormalization group approaches~\cite{StueckelbergdeBreidenbach:1952pwl, Peterman:1978tb}. Following the standard procedures of PMCs~\cite{Shen:2017pdu, Yan:2022foz}, all the RGE-involved non-conformal terms of the above conventional series (\ref{rhoConv2}) of $R_{4}(\mu_r)$ shall be removed from the series and be adopted for fixing the correct magnitude of $\alpha_s$ of the process, one then obtains a scale-invariant conformal series. Up to five-loop level, we have
\begin{align}
	\label{R}
	R_{4}|_{\rm{PMCs}}=\sum_{i=1}^{4}\hat{r}_{i,0}a^i(Q_*)+ {\cal O}(a^5),
\end{align}
where $Q_*$ is the PMC scale, which can be determined by the following equation
\begin{widetext}
\begin{eqnarray}
\ln\frac{Q_{*, {\rm N^{\it l}LL}}^2}{Q^2} && =  -\frac{\sum_{k=1}^{l+2}\sum_{i=1}^{l-k+2} \left[ (-1)^i\Delta_{n,k}^{(i-1)}\hat{r}_{k+i,i}(n+k-1) a^k(Q_{*, {\rm N^{\it l}LL}})\right]} {\sum_{\eta=1}^{l+1} \sum_{k=1}^{l+2} \sum_{i=\eta}^{l-k+2}\left[(-1)^{i} (n+k-1)  C_i^\eta\Delta_{n,k}^{(i-1)}\hat{r}_{k+i-\eta,i-\eta} L^{\eta-1}_{Q_{*, {\rm N^{\it l-1}LL}}} a^k(Q_{*, {\rm N^{\it l}LL}})\right]}  \\
&& = \sum_{i=0}^{2}S_ia^i(Q_{*, {\rm N^{\it l}LL}}) \label{fullexpQ}
\end{eqnarray}
\end{widetext}
where $L_{Q_{*, {\rm N^{\it l-1}LL}}}=\ln{Q_{*, {\rm N^{\it l-1}LL}}^2}/{Q^2}$. In the second line, e.g. Eq.(\ref{fullexpQ}), we have expanded the series in the nominator and denominator as power series over $a=\alpha_s/\pi$; and their precision depend on how many loop terms for the pQCD approximant $R_{n}$ have been known. That is, by using $R_{2}$, $R_{3}$ and $R_{4}$ accordingly, the PMC scale shall be fixed at the LL-accuracy, NLL-accuracy and N$^2$LL-accuracy, respectively. More explicitly, up to N$^4$LO level, we only need to know the first three functions $\Delta_{n,k}^{(0,1,2)}$, which are
\begin{align}
	&\Delta_{n,k}^{(0)}=1 , \notag \\
	&\Delta_{n,k}^{(1)}=-\frac{1}{2}\sum_{i=0}^{+\infty}(n+k+i)\beta_ia^{i+1} , \notag \\
	&\Delta_{n,k}^{(2)}=\frac{1}{3!}\sum_{i=0}^{+\infty}\sum_{j=0}^{+\infty}(n+k+i)(n+i+j+k+1) \notag\\
    &\quad\quad\quad\quad\quad\quad\quad\quad \times\beta_i\beta_ja^{i+j+2}.
\end{align}
And the functions $S_i$ with $i=(0,1,2)$ that are defined in the second line (\ref{fullexpQ}) are
 \begin{align}
 	&S_0=-\frac{\hat{r}_{2,1}}{\hat{r}_{1,0}}\\
 &S_1=\frac{2(\hat{r}_{2,0}\hat{r}_{2,1}-\hat{r}_{1,0}\hat{r}_{3,1})}{\hat{r}_{1,0}^2}+\frac{(\hat{r}_{2,1}^2-\hat{r}_{1,0}\hat{r}_{3,2})}{\hat{r}_{1,0}^2}\beta_0\\
 &S_2=\frac{4(\hat{r}_{1,0}\hat{r}_{2,0}\hat{r}_{3,1}-\hat{r}_{2,0}^2\hat{r}_{2,1})+3(\hat{r}_{1,0}\hat{r}_{2,1}\hat{r}_{3,0}-\hat{r}_{1,0}^2\hat{r}_{4,1})}{\hat{r}_{1,0}^3}\notag\\&+\frac{3(\hat{r}_{2,1}^2-\hat{r}_{1,0}\hat{r}_{3,2})}{2\hat{r}_{1,0}^2}\beta_1-\bigg[\frac{2\hat{r}_{2,0}^2\hat{r}_{2,1}-\hat{r}_{1,0}\hat{r}_{2,0}(6\hat{r}_{3,1}+2\hat{r}_{3,2})}{\hat{r}_{1,0}^3}\notag\\&-\frac{3(\hat{r}_{2,0}\hat{r}_{2,1}^2+\hat{r}_{1,0}^2\hat{r}_{4,2})}{\hat{r}_{1,0}^3}\bigg]\beta_0+\bigg[\frac{(\hat{r}_{1,0}\hat{r}_{2,0}\hat{r}_{3,2}-\hat{r}_{1,0}^2\hat{r}_{4,3})}{\hat{r}_{1,0}^3}\notag\\&+\frac{2(\hat{r}_{1,0}\hat{r}_{2,0}\hat{r}_{3,2}-\hat{r}_{2,1}^3)}{\hat{r}_{1,0}^3}\bigg]\beta_0^2
\end{align}

The predictive power of pQCD prediction also depends on a reliable way of estimating the contributions of unknown higher-order (UHO) terms. The Bayesian-based approach provides such a way of estimating the UHO contribution, which predicts the magnitude of the UHO-terms based on an optimized analysis of probability density distribution. The Bayesian analysis constructs probability distributions in which Bayes' theorem is used to iteratively update the probability as new information becomes available~\cite{Cacciari:2011ze, Bagnaschi:2014wea, Bonvini:2020xeo, Duhr:2021mfd, Shen:2022nyr}. The interested reader may turn to Ref.\cite{Shen:2022nyr} to know the recent progresses on the Bayesian analysis. We put the key formulas in the following for self-consistency.

If the perturbative approximant starts at the initial order $O(\alpha_s^l)$ and stops at the $k_{\rm th}$ order $O(\alpha_s^k)$, the corresponding perturbatively calculable physical observable can be schematically represented as
\begin{align}
	\rho_k=\sum_{i=l}^{k}c_i\alpha_s^i ,
\end{align}
where $c_i$ are expansion coefficients. Doing the replacing $\rho_k\to R_{n}$, $l\to 1$ and $c_i\to r_i$ ($\hat{r}_{i,0}$) in the following formulas, we get the required formulas for the conventional (PMC) series of $R_{n}$. By taking three reasonable hypotheses, we obtain the probability density function (p.d.f) for the unknown higher-order coefficient $c_n$,
\begin{align}
	\label{fc}
	f_c(c_n|c_l,...,c_k)=\begin{cases}
		\frac{n_c}{2(n_c+1)\bar{c}_{(k)}},|c_n|\leq\bar{c}_{(k)}\\
		\frac{n_c\bar{c}^{n_c}_{(k)}}{2(n_c+1)|c_n|^{n_c+1}},|c_n|>\bar{c}_{(k)}
	\end{cases}.
\end{align}
where $\bar{c}_{(k)}$= Max$\{|c_l|,...,|c_k|\}$, and $n_c=k-l+1$, which represents the number of known perturbative coefficients, $c_l,...,c_k$. Using Eq.(\ref{fc}), one then derives the conditional p.d.f. for the uncalculated higher-order term $\delta_n=c_n\alpha_s^n$, $(n>k)$. Especially for the one-order higher UHO-term with $n=k+1$, the conditional p.d.f. of $\delta_{k+1}$ and $\rho_{k+1}$ with given coefficients $c_l,...,c_k$, denoted by $f_\delta(\delta_{k+1}|c_l,...c_k)$ and $f_\rho(\rho_{k+1}|c_l,...c_k)$, respectively, read
\begin{widetext}
\begin{eqnarray}\label{fdelta}
f_\delta(\delta_{k+1}|c_l,\dots,c_k) = \left(\frac{n_c}{n_c+1}\right)\frac{1}{2\alpha_s^{k+1} \bar{c}_{(k)}}
\left\{
\begin{array}{ll}
1, & |\delta_{k+1}|\leq \alpha_s^{k+1}\bar{c}_{(k)}\\[8pt]
\left(\frac{\alpha_s^{k+1}\bar{c}_{(k)}}{|\delta_{k+1}|}\right)^{n_c+1}, & |\delta_{k+1}|>\alpha_s^{k+1}\bar{c}_{(k)}
\end{array}
\right. ,
\end{eqnarray}
\begin{eqnarray}
\label{frho}
f_\rho(\rho_{k+1}|c_l,\cdots,c_k) = \left(\frac{n_c}{n_c+1}\right)\frac{1}{2\alpha_s^{k+1} \bar{c}_{(k)}}
\left\{
\begin{array}{ll}
1, & |\rho_{k+1}-\rho_{k}|\leq \alpha_s^{k+1}\bar{c}_{(k)}\\[8pt]
\left(\frac{\alpha_s^{k+1}\bar{c}_{(k)}}{|\rho_{k+1}-\rho_{k}|}\right)^{n_c+1}, & |\rho_{k+1}-\rho_{k}|>\alpha_s^{k+1}\bar{c}_{(k)}
\end{array}
\right. \, .
\end{eqnarray}
\end{widetext}

One usually estimates the central value of $\rho_{k+1}$ to be its expectation value $E(\rho_{k+1})$ and takes its uncertainty as its standard deviation, $\sigma_{k+1}$. The expectation value $E(\rho_{k+1})$ can be related to the expectation value of $\delta_{k+1}$, i.e. $E(\rho_{k+1})=E(\delta_{k+1})+\rho_k$. For the present prior distribution, $E(\delta_{k+1})=0$, due to the fact that the symmetric probability distribution (\ref{fdelta}) is centered at zero. To predict the magnitude of $\delta_{k+1}$ consistently, it is useful to define a critical degree-of-belief (DoB), $p_c\%$, which equals to the least value of $p\%$ that satisfies the following equations,
\begin{eqnarray}
\hspace{-5mm} \rho_{i-1}+c_{i}^{(p)}\alpha_s^{i}\geq \rho_{i}+c_{i+1}^{(p)}\alpha_s^{i+1}, \; (i=l+1,\cdots,k), \\
\hspace{-5mm} \rho_{i-1}-c_{i}^{(p)}\alpha_s^{i}\leq \rho_{i}-c_{i+1}^{(p)}\alpha_s^{i+1}, \; (i=l+1,\cdots,k).
\end{eqnarray}
Thus, for any $p\%\geq p_c\%$, the error bars determined by the $p\%$-credible intervals (CIs) provide consistent estimates for the magnitude of $\delta_{k+1}$. The value of $p_c\%$ is nondecreasing when $k$ increases. Practically, we will adopt the smallest $p_s\%$-CI so as to obtain a consistent and high DoB estimation, i.e.
\begin{eqnarray}
[E(\rho_{k+1})-c_{k+1}^{(p_s)}\alpha_s^{k+1},E(\rho_{k+1})+c_{k+1}^{(p_s)}\alpha_s^{k+1}],
\end{eqnarray}
as final estimate for $\rho_{k+1}$, where $p_s\% ={\rm Max}\{p_c\%, p_\sigma\% \}$. Here $p_\sigma\%$ represents the DoB for the $1\sigma$-interval, and $\rho_{k+1}\in[E(\rho_{k+1})-\sigma_{k+1},E(\rho_{k+1})+\sigma_{k+1}]$.

\section{Numerical results and discussions}

To do the numerical calculation, we take the values of input parameters from Particle Data Group~\cite{ParticleDataGroup:2022pth}, e.g. the $W$-boson mass $M_W$ = 80.377 GeV, the $\tau$-lepton mass $M_\tau$ = 1.7769 GeV, the $b$-quark pole mass $M_b$ = 4.78 GeV, the $c$-quark pole mass $M_c$ = 1.67 GeV, the $t$-quark pole mass $M_t$ = 172.69 GeV, and the Higgs mass $M_H$ = 125.25 GeV. The Fermi constant $G_F = 1.1664\times10^{-5}~\rm{GeV}^{-2}$ and the fine structure constant $\alpha$ = 1/137.036. We have assumed the running of $\alpha_s$ at the four-loop level, the QCD asymptotic scale $\Lambda_{\rm QCD}$ is determined by using $\alpha_s(M_Z)= 0.1179$, which gives $\Lambda^{n_f=5}_{\rm QCD}=0.2072$ GeV.

For the process $H\to\gamma\gamma$, its QCD correction $R_n$ under the $\overline{\rm MS}$-scheme has been calculated up to N$^{4}$LO level. The initial fixed-order pQCD series is scheme and scale dependent~\footnote{A way of achieving scheme-and-scale invariant prediction directly from the initial series, which is called as principal of minimum sensitivity (PMS)~\cite{Stevenson:1980du, Stevenson:1981vj} has been suggested in the literature. It assumes that all uncalculated higher-order terms give zero contribution and determines the optimal scheme and scale by requiring the slope of the pQCD series over scheme and scale choices vanish. Since the PMS breaks the standard renormalization group invariance~\cite{Wu:2014iba}, it cannot be treated as a strict solution of conventional scheme-and-scale ambiguities, which however could be treated as an effective treatment~\cite{Ma:2014oba, Ma:2017xef}.}. As has been discussed above, after applying the PMC, the resultant conformal series becomes scheme-and-scale invariant. We present the scale-invariant conformal coefficients $\hat{r}_{i,0} (i=1,\cdots,4)$ in Table~\ref{rknown}, where the scale-dependent coefficients $r_i$ at $\mu_r=M_H/2$, $M_H$ and $2M_H$ are also presented as comparisons.

\begin{table}[htb]
\centering
\caption{The $\overline{\rm MS}$ coefficients $\hat{r}_{i,0}$ and $r_i$ for $R_4$. The coefficients $r_i$ are also scale dependent and their values under three typical scale choices, e.g. $\mu_r=M_H/2$ $M_H$ and $2M_H$, are given as comparisons.}
\label{rknown}
\begin{tabular}{cccccccc}
		\hline
		~~~&~~~ $i=1$~~~&~~~$i=2$~~~&~~~$i=3$~~~&~~~$i=4$~~~ \\
			\hline		
		$r_i(\mu_{r}=M_H/2)$          & $1.4070$ & $-0.9874 $ & $-0.4084 $ & $3.3437$ \\
		\hline		
		$r_i(\mu_{r}=M_H)$          & $1.4070$ & $0.2024 $ & $-1.6545 $ & $-0.3693$ \\
			\hline		
		$r_i(\mu_{r}=2M_H)$          & $1.4070$ & $1.5282 $ & $-0.3456 $ & $-2.4065$ \\
		\hline
		$\hat{r}_{i,0}$   & $1.4070$ & $1.3387 $ & $ -3.6304 $ &  $4.5695$  \\
		\hline		
\end{tabular}
\end{table}

Using the expansion coefficients of the QCD corrections $R_{2}$, $R_{3}$ and $R_{4}$, the PMC scale can be fixed at the LL-accuracy, NLL-accuracy and N$^2$LL-accuracy, respectively. And we obtain
\begin{eqnarray}
  Q_{*,{\rm LL}}  &=& 242.791 ~{\rm GeV},  \\
  Q_{*,{\rm NLL}} &=& 193.457 ~{\rm GeV},  \\
  Q_{*,{\rm N^{2}LL}} &=& 213.603  ~{\rm GeV}.
\end{eqnarray}
The $|Q_{*,{\rm N^{2}LL}}-Q_{*,{\rm NLL}}| < |Q_{*,{\rm NLL}}-Q_{*,{\rm LL}}|$ indicates that the expansion series of $\ln{Q_{*}^2}/{Q^2}$ has perturbative nature. Together with the fact that its higher-order terms will suffer from both $\alpha_s$-power suppression and exponential suppression, thus the residual scale dependence of $Q_{*}$ due to even higher-order terms of $R_{n}$ will be highly suppressed, whose effects to the magnitude of $\alpha_s$ is negligible. The PMC predictions of $R_{2}$, $R_{3}$ and $R_{4}$ are
\begin{eqnarray}
	R_{2}|_{\rm PMC} &=& 0.159493 ~\rm{keV}, \\
	R_{3}|_{\rm PMC} &=& 0.159969 ~\rm{keV}, \\
	R_{4}|_{\rm PMC} &=& 0.158517 ~\rm{keV}.
\end{eqnarray}

\begin{table*}[htb]
\centering
\caption{The N$^4$LO QCD corrections $R_4=\sum_{i=1}^{4}\Delta_i$ of $\Gamma(H\to\gamma\gamma)$ under conventional (Conv.) and PMC scale-settings, respectively. $\Delta_i$ represents individual decay width at NLO-, N$^{2}$LO-, N$^{3}$LO- or N$^{4}$LO- level, respectively. Three typical values $\mu_{r}=M_H/2$, $M_H$ and $2M_H$ are adopted to show renormalization scale uncertainty. }
\label{Rtotal}
\begin{tabular}{cccccccc}
\hline
~~~~&~~ ~~&~~$i={1}$~~&~~$i={2}$~~&~~$i={3}$~~&~~$i={4}$~~&~~$R_4(\mu_r)$~~ \\
\hline
	& $\mu_r=M_H/2$              & $0.17589$ & $-0.01543$ & $-0.00080$ & $ 0.00082$ & $0.16048$   \\
${\Delta_i \rm{(KeV)}}|_{\rm Conv.}$      & $\mu_r=M_H$    & $0.15830$ & $ 0.00256$ & $ -0.00236$ & $-0.00006$ & $0.15845$   \\
	& $\mu_r=2M_H$              &$ 0.14467 $ & $ 0.01616$ &  $-0.00038$ & $-0.00027$ & $0.16018$  \\
\hline
${\Delta_i\rm{(KeV)}}|_{\rm PMC}$  & $\mu_r\in[Q/2,2Q]$  & $0.14744$ & $ 0.01470$ &  $-0.00418$ & $ 0.00055$ & $0.15852$  \\
\hline
\end{tabular}
\end{table*}

Table~\ref{Rtotal} shows that the N$^4$LO QCD corrections $R_4=\sum_{i=1}^{4}\Delta_i$ under conventional and PMC scale-settings, where $\Delta_i$ represents the individual decay width at the NLO-, the N$^{2}$LO-, the N$^{3}$LO- or the N$^{4}$LO- level, respectively. Three typical scales $\mu_{r}=M_H/2$, $M_H$ and $2M_H$ are adopted to show the conventional renormalization scale uncertainty. Table~\ref{Rtotal} shows that under conventional scale-setting, the separate decay widths $\Delta_i$ are highly scale dependent, and due to the large cancellation among different orders, the net scale dependence of the N$^4$LO prediction $R_4$ becomes small $\sim 12.8\%$ for $\mu_{r}\in [M_H/2, 2M_H]$. After applying the PMC, both $\Delta_i$ and $R_4$ are scale independent. This confirms the observation that if the correct magnitude of $\alpha_s$ of a pQCD series has been determined by using the RG-involved $\{\beta_i\}$-terms, indicating well matching of $\alpha_s$ with its expansion coefficients, one will achieve a precise scale independent pQCD prediction. Such scale independent nature of the pQCD approximant can be treated as its intrinsic perturbative property. Due to good perturbative nature of the PMC series of $R_n$, the difference between the magnitudes of $R_n$ and $R_{n-1}$ becomes smaller with the increment of the given loop numbers.

\begin{table*}[htb]
\centering
\caption{The predicted smallest 95.5$\%$ CIs for the scale-dependent conventional coefficients $r_i(\mu_{r})$ at the scale $\mu_{r}=M_H$ and the scale-invariant coefficients $\hat{r}_{i,0}(i=3, 4, 5)$ of $R_n(\mu_r=M_H)$ via the Bayesian approach, where $M_H=125.25~{\rm GeV}$. The values from given series (``ECs") are presented as comparisons.}
\label{rBayesian}
\begin{tabular}{cccccccc}
		\hline
		~~~~&~~~ $r_2(M_H)$~~~&~~~$r_3(M_H)$~~~&~~~$r_4(M_H)$~~~&~~~$r_5(M_H)$~~~ \\
		\hline		
		CI          & $[-15.6334,15.6334]$ & $[-3.8294,3.8294] $ & $[-2.9303,2.9303] $ & $[-2.4023,2.4023]$ \\
		EC          & $0.2024$ & $-1.6545$ & $-0.3693$ & $-$ \\
		\hline
		~~~~&~~~ $\hat{r}_{2,0}$~~~&~~~$\hat{r}_{3,0}$~~~&~~~$\hat{r}_{4,0}$~~~&~~~$\hat{r}_{5,0}$~~~ \\
		\hline		
		CI          & $[-15.6334,15.6334]$ & $[-3.8294,3.8294]$ & $[-6.4298,6.4298]$ & $[-6.6348,6.6348]$ \\
		EC          & $1.3387$ & $-3.6304$ & $4.5695$ & $-$ \\
		\hline	
\end{tabular}
\end{table*}

Under Bayesian approach, we will predict the magnitude of the unknown coefficient $c_{i+1}$ from the known ones $\{c_1,\cdots,c_i\}$ with $c_i\to r_i$ ($\hat{r}_{i,0}$) for conventional (PMC) series, respectively. Our results are listed in Table~\ref{rBayesian}. From Tables~\ref{rknown}, \ref{rBayesian}, we can see that the exact values of $r_{i,0}(i=2, 3, 4,)$ and $r_i(i=2, 3, 4)$ lie within the predicted 95.5$\%$ CIs. Moreover, we can obtain the
smallest 95.5$\%$ credible intervals (CIs) for the perturbative coefficients $r_5(\mu_r=M_H)$ and $r_{5,0}$, which are $r_5\in[-2.4023,2.4023]$ and $r_{5,0}\in[-6.6348,6.6348]$, respectively. The values from given series (``ECs") are presented as comparisons.

Using the estimated $r_5(M_H)$ and $\hat{r}_{5,0}$, the error of $\Gamma_{\rm{H}}$ caused by the UHO-terms for conventional series and PMC series under the Bayesian approach (B.A.) are
\begin{align}
	&\Delta\Gamma_H|_{\rm{Conv.}}^{\rm{UHO}}=\pm8.523\times10^{-5}~{\rm keV}, \label{deltaconvUHO}\\
	&\Delta\Gamma_H|_{\rm{PMC}}^{\rm{UHO}}=\pm1.65\times10^{-4}~{\rm keV}. \label{deltapmcsUHO}
\end{align}
By further taking $\mu_r\in[M_H/2,2M_H]$, the conventional series also has the following scale uncertainty
\begin{align}
	\label{deltamur}
\Delta\Gamma_H|^{\mu_r}_{\rm{Conv.}}=(^{+2.03\times10^{-3}}_{-1.02\times10^{-5}})~{\rm keV}.
\end{align}
Then as a combination, the net errors caused by the N$^5$LO UHO-terms in conventional and PMC series are
\begin{align}
	&\Delta\Gamma_H|_{\rm{Conv.}}=(^{+2.03\times10^{-3}}_{-8.58\times10^{-5}})~{\rm keV}\\
	&\Delta\Gamma_H|_{\rm{PMC}}=\pm1.65\times10^{-4}~{\rm keV}\label{deltapmcs}
\end{align}
where $\mu_r\in[M_H/2,2M_H]$.

\begin{figure}[htb]
	\centering
	\includegraphics[width=0.45\textwidth]{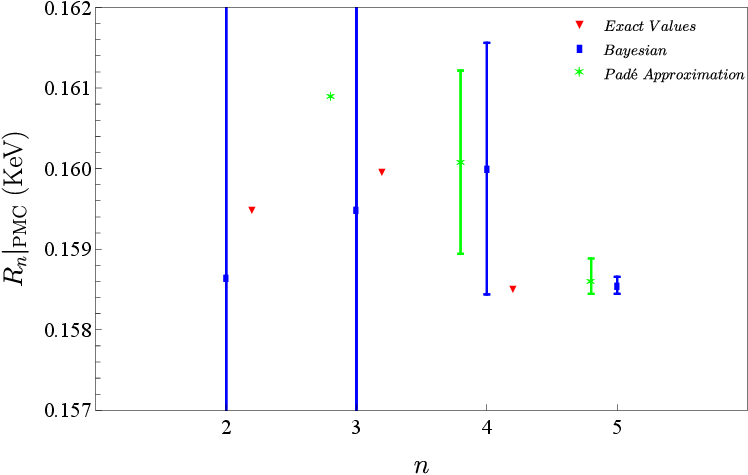}
	\caption{The predicted values for the pQCD correction $R_n|_{\rm PMC}$ under the  Pad$\acute{\rm{e}}$ approximation approach (PAA) and Bayesian approach (B.A.) at different orders, respectively. The blue rectangles together with the error bars, are for B.A., the green error bars are brought by different types of PAA, and the exact values of the $R_n(M_H)|_{\rm PMC}$ at different orders, respectively.}
	\label{fig:PAABAcompare3}
\end{figure}

In addition, for the more precise PMC series, we also adopt another usual way of estimating UHO contributions, e.g. the Pad$\acute{\rm{e}}$ approximation approach (PAA)~\cite{Basdevant:1972fe, Samuel:1992qg, Samuel:1995jc} to estimate the UHO-terms of $R_{\rm{n}}$. The PAA works when we have known enough higher orders, e.g. $n\geq 2$ for the present case. The PAA has an intrinsic error due to the existence of different types of generating function~\cite{Du:2018dma}, and we will take the result of $[0/n-1]$-type as its central value and the results of other types are treated as its uncertainty. More explicitly, to estimate the N$^3$LO magnitude from the given N$^2$LO series, we have $[0/1]$-type generating function; to estimate the N$^4$LO magnitude from the given N$^3$LO series, we have $[0/2]$-type and $[1/1]$-type generating functions; to estimate the N$^5$LO magnitude from the given N$^4$LO series, we have $[0/3]$-type, $[1/2]$-type and $[2/1]$-type generating functions; and etc. We put the results in Fig.(\ref{fig:PAABAcompare3}), where the ``Exact Values" together with the Bayesian approach (B.A.) and Pad$\acute{\rm{e}}$ approximation approach (PAA). ones are presented. Fig.(\ref{fig:PAABAcompare3}) shows that for the B.A. approach, the ``exact" value are always within the predicted error band, the predicted one-order higher UHO error band is always within the predicted one-order lower UHO error band, and the predicted UHO values become more accurate when more loop terms have been known. Thus if one has enough higher-order information to tame the probability density function, one may achieve precise contribution of the UHO terms. For the PAA, the ``exact" N$^4$LO value is outside of the predicted error bar, and the predicted N$^5$LO error bar becomes better and is consistent with the B.A. one. In this sense, at least for the present case, the B.A. approach is more effective than PAA.

From Eq.(\ref{GammaH}), there are other error sources such as $\Delta M_H$, $\Delta m_t$ and $\Delta\alpha_s(M_Z)$ for the total decay width $\Gamma(H\to\gamma\gamma)$. For the purpose, we take $\Delta M_H=\pm0.17$ GeV, $\Delta m_t=\pm0.30$ GeV and $\Delta \alpha_s(M_z)=\pm0.0009$ GeV~\cite{ParticleDataGroup:2022pth} to show their effects. When discussing the error caused by one parameter, the other parameters will be fixed as their center values. And we have
\begin{align}
	&\Delta\Gamma_H|_{\rm Conv.}^{\Delta M_H}=(^{+5.455\times10^{-2}}_{-5.423\times10^{-2}})~{\rm keV}, \\
		&\Delta\Gamma_H|_{\rm PMC}^{\Delta M_H}=(^{+5.453\times10^{-2}}_{-5.421\times10^{-2}})~{\rm keV}, \\
	&\Delta\Gamma_H|_{\rm Conv.}^{\Delta m_t}=(^{+6.999\times10^{-4}}_{-7.040\times10^{-4}})~{\rm keV}, \\
	&\Delta\Gamma_H|_{\rm PMC}^{\Delta m_t}=(^{+7.004\times10^{-4}}_{-7.045\times10^{-4}})~{\rm keV}, \\
	&\Delta\Gamma_H|_{\rm Conv.}^{\Delta \alpha_s(M_Z)}=(^{+1.071\times10^{-3}}_{-1.072\times10^{-3}})~{\rm keV}, \\
	&\Delta\Gamma_H|_{\rm PMC}^{\Delta \alpha_s(M_Z)}=(^{+1.061\times10^{-3}}_{-1.062\times10^{-3}})~{\rm keV}.
\end{align}

By adding all the mentioned errors in quadrature, our final results for the total decay $\Gamma_H$ of $H\to\gamma\gamma$ using the B.A. approach are
\begin{align}
	&\Gamma_H|^{\rm B.A.}_{\rm Conv.}=9.56497^{+0.05461}_{-0.05424}~{\rm keV}, \\
	&\Gamma_H|^{\rm B.A.}_{\rm PMC}=9.56504^{+0.05455}_{-0.05422}~{\rm keV}
\end{align}
whose net errors are $1.138\%$ and $1.137\%$. This shows that since the QCD correction has been calculated up to N$^4$LO level, the main errors are dominated by $\Delta M_{H}$.

\subsection*{The fiducial cross section of $\sigma_{\rm fid}(pp\to H\to\gamma\gamma)$}

\begin{figure}[htb]
\centering
\includegraphics[width=0.5\textwidth]{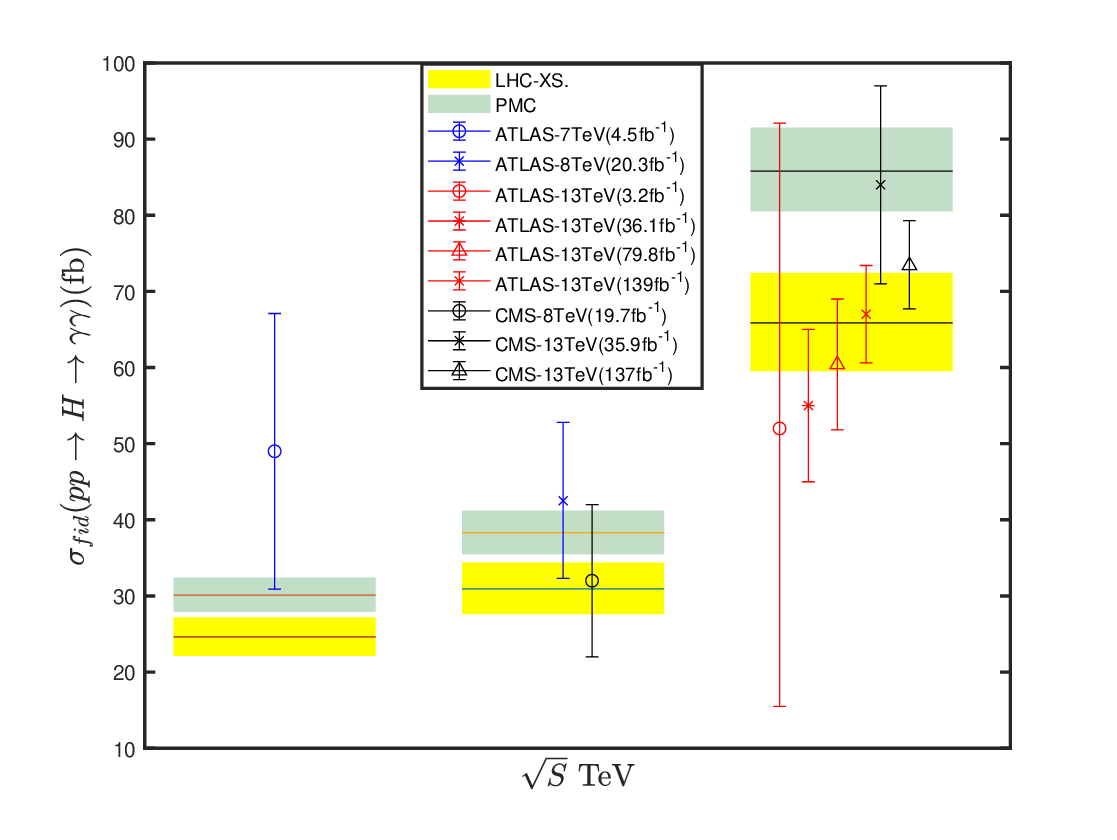}
\caption{The fiducial cross section $\sigma_{\rm fid}(pp\rightarrow H\rightarrow\gamma\gamma)$ using the $\Gamma(H\rightarrow\gamma\gamma)$ up to N$^4$LO level. The LHC-XS prediction, the ATLAS measurements~\cite{TheATLAScollaboration:2015poe, ATLAS:2017myr, Alves:2022cuz, ATLAS:2022jsi} and the CMS measurement\cite{CMS:2017nyv, CMS:2015qgt, CMS:2022wpo} are presented as a comparison.}
\label{fig:shiyan1}
\end{figure}

As an application of $H\to\gamma\gamma$ decay width, we estimate the ``fiducial cross section" of the process $pp\to H\to\gamma\gamma$. The fiducial cross-section $\sigma_{\rm fid}$ can be written as
\begin{align}
	\sigma_{\rm fid}(pp\to H\to\gamma\gamma)=\sigma_{\rm Incl} {\cal B}_{H\to\gamma\gamma} A
\end{align}
where $A$ is the acceptance factor, whose value for different collision energies can find in Ref.\cite{TheATLAScollaboration:2015poe}. The $B_{H\to\gamma\gamma}$ represents the branching ratio of $H\to\gamma\gamma$. By using the $\Gamma(H\to\gamma\gamma)$ with conventional scale-setting approach, the LHC-XS group gives ${\cal B}_{H\rightarrow\gamma\gamma}=0.00227^{+0.00206}_{-0.00208}$~\cite{LHCHiggsCrossSectionWorkingGroup:2016ypw}. The inclusive cross-section $\sigma_{\rm Incl}$ predicted by LHC-XS group is given in Ref.\cite{LHCHiggsCrossSectionWorkingGroup:2013rie}. The results are $\sigma_{\rm{fid}}(pp\to H\to\gamma\gamma)|_{\rm LHC-XS}=24.63^{+2.55}_{-2.50}$ fb, $30.93^{+3.44}_{-3.33}$ fb, and $65.86^{+6.58}_{-6.33}$ fb for the proton-proton center-of-mass collision energy $\sqrt{\rm S}$= 7, 8 and 13 TeV, respectively, which has been measured by ATLAS and CMS collaborations with increasing integrated luminocities~\cite{TheATLAScollaboration:2015poe, ATLAS:2017myr, ATLAS:2022jsi, Alves:2022cuz, CMS:2017nyv, CMS:2015qgt, CMS:2022wpo}. Taking the same inputs as those of Refs.\cite{LHCHiggsCrossSectionWorkingGroup:2013rie, Wang:2016wgw, LHCHiggsCrossSectionWorkingGroup:2016ypw}, e.g. $M_H$=125 GeV and $M_t$=173.3 GeV, and using the QCD corrections up to N$^4$LO level, we obtain $\sigma_{\rm{fid}}(pp\to H\to \gamma\gamma)|_{\rm PMC}=30.1^{+2.3}_{-2.2}$ fb, $38.3^{+2.9}_{-2.8}$ fb, and $85.5^{+5.7}_{-5.3}$ fb for the proton-proton center-of-mass collision energy $\sqrt{\rm S}$= 7, 8 and 13 TeV, respectively. As an intuitive comparison of the experimental data and theoretical results, we present the results in Fig.(\ref{fig:shiyan1}). It shows that when $\sqrt{\rm S}$=7, 8 TeV, the theoretical results are consistent with the experimental measurements; and when $\sqrt{\rm S}$=13 TeV, the measured values of ATLAS and CMS differ significantly, and the theoretical results are closer to the data of CMS.

\section{Summary}

By using the PMC scale-setting approaches, all non-conformal terms have been adopted to set the correct magnitude of $\alpha_s$ with the help of RGE, and the resultant pQCD series becomes more precise without conventional scheme-and-scale independence. In this paper, we have calculated the decay width $\Gamma(H\rightarrow\gamma\gamma)$ up to N$^4$LO QCD corrections. The Bayesian approach has been applied to estimate the uncalculated N$^5$LO contribution, which is only about $\pm1.65\times10^{-4}$ keV for the case of smallest $95.5\%$ credible interval. After taking all the mentioned errors into consideration, we predict $\Gamma_H|^{\rm B.A.}_{\rm PMC}=9.56504^{+0.05455}_{-0.05422}$ keV. Thus by using the Bayesian approach, one can consistently obtain high reliability estimations of UHO-contributions by using convergent and scale-independent PMC series, greatly improving the prediction ability of pQCD.

\hspace{2cm}

\noindent {\bf Acknowledgments:} The authors would like to thank Qing, Yu for helpful discussions. This work was supported in part by the Chongqing Graduate Research and Innovation Foundation under Grant No. CYB23011 and No.ydstd1912, and by the Natural Science Foundation of China under Grant No.12175025 and No.12147102.


\begin{thebibliography}{99}	

\bibitem{ATLAS:2012yve}
G.~Aad \textit{et al.} [ATLAS],
``Observation of a new particle in the search for the Standard Model Higgs boson with the ATLAS detector at the LHC,''
Phys. Lett. B \textbf{716}, 1 (2012).

\bibitem{CMS:2012qbp}
S.~Chatrchyan \textit{et al.} [CMS],
``Observation of a New Boson at a Mass of 125 GeV with the CMS Experiment at the LHC,''
Phys. Lett. B \textbf{716}, 30 (2012).

\bibitem{ILC:2013jhg}
H.~Baer \textit{et al.} [ILC],
``The International Linear Collider Technical Design Report - Volume 2: Physics,''
[arXiv:1306.6352 [hep-ph]].

\bibitem{CEPCStudyGroup:2018ghi}
J.~B.~Guimar\~aes da Costa \textit{et al.} [CEPC Study Group],
``CEPC Conceptual Design Report: Volume 2 - Physics \& Detector,''
[arXiv:1811.10545 [hep-ex]].

\bibitem{FCC:2018byv}
A.~Abada \textit{et al.} [FCC],
``FCC Physics Opportunities: Future Circular Collider Conceptual Design Report Volume 1,''
Eur. Phys. J. C \textbf{79}, 474 (2019).

\bibitem{ParticleDataGroup:2022pth}
R.~L.~Workman \textit{et al.} [Particle Data Group],
``Review of Particle Physics,''
PTEP \textbf{2022}, 083C01 (2022).

\bibitem{Yu:2018hgw}
Q.~Yu, X.~G.~Wu, S.~Q.~Wang, X.~D.~Huang, J.~M.~Shen and J.~Zeng,
``Properties of the decay $H\to\gamma\gamma$ using the approximate $\alpha_s^4$ corrections and the principle of maximum conformality,''
Chin. Phys. C \textbf{43}, 093102 (2019).

\bibitem{Ellis:1975ap}
J.~R.~Ellis, M.~K.~Gaillard and D.~V.~Nanopoulos,
``A Phenomenological Profile of the Higgs Boson,''
Nucl. Phys. B \textbf{106}, 292 (1976).

\bibitem{Shifman:1979eb}
M.~A.~Shifman, A.~I.~Vainshtein, M.~B.~Voloshin and V.~I.~Zakharov,
``Low-Energy Theorems for Higgs Boson Couplings to Photons,''
Sov. J. Nucl. Phys. \textbf{30}, 711 (1979).

\bibitem{Zheng:1990qa}
H.~Q.~Zheng and D.~D.~Wu,
``First order QCD corrections to the decay of the Higgs boson into two photons,''
Phys. Rev. D \textbf{42}, 3760 (1990).

\bibitem{Dawson:1992cy}
S.~Dawson and R.~P.~Kauffman,
``QCD corrections to H ---\ensuremath{>} gamma gamma,''
Phys. Rev. D \textbf{47}, 1264 (1993).

\bibitem{Djouadi:1990aj}
A.~Djouadi, M.~Spira, J.~J.~van der Bij and P.~M.~Zerwas,
``QCD corrections to gamma gamma decays of Higgs particles in the intermediate mass range,''
Phys. Lett. B \textbf{257}, 187 (1991).

\bibitem{Djouadi:1993ji}
A.~Djouadi, M.~Spira and P.~M.~Zerwas,
``Two photon decay widths of Higgs particles,''
Phys. Lett. B \textbf{311}, 255 (1993).

\bibitem{Melnikov:1993tj}
K.~Melnikov and O.~I.~Yakovlev,
``Higgs ---\ensuremath{>} two photon decay: QCD radiative correction,''
Phys. Lett. B \textbf{312}, 179 (1993).

\bibitem{Inoue:1994jq}
M.~Inoue, R.~Najima, T.~Oka and J.~Saito,
``QCD corrections to two photon decay of the Higgs boson and its reverse process,''
Mod. Phys. Lett. A \textbf{9}, 1189 (1994).

\bibitem{Spira:1995rr}
M.~Spira, A.~Djouadi, D.~Graudenz and P.~M.~Zerwas,
``Higgs boson production at the LHC,''
Nucl. Phys. B \textbf{453}, 17 (1995).

\bibitem{Fleischer:2004vb}
J.~Fleischer, O.~V.~Tarasov and V.~O.~Tarasov,
``Analytical result for the two loop QCD correction to the decay H ---\ensuremath{>} 2 gamma,''
Phys. Lett. B \textbf{584}, 294 (2004).

\bibitem{Harlander:2005rq}
R.~Harlander and P.~Kant,
``Higgs production and decay: Analytic results at next-to-leading order QCD,''
JHEP \textbf{12}, 015 (2005).

\bibitem{Anastasiou:2012hx}
C.~Anastasiou, S.~Buehler, F.~Herzog and A.~Lazopoulos,
``Inclusive Higgs boson cross-section for the LHC at 8 TeV,''
JHEP \textbf{04}, 004 (2012).

\bibitem{Maierhofer:2012vv}
P.~Maierh\"ofer and P.~Marquard,
``Complete three-loop QCD corrections to the decay H -\ensuremath{>} \textbackslash{}gamma \textbackslash{}gamma,''
Phys. Lett. B \textbf{721}, 131 (2013).

\bibitem{Sturm:2014nva}
C.~Sturm,
``Higher order QCD results for the fermionic contributions of the Higgs-boson decay into two photons and the decoupling function for the $\overline{\text{ MS }}$ renormalized fine-structure constant,''
Eur. Phys. J. C \textbf{74}, 2978 (2014).

\bibitem{Marquard:2016dcn}
P.~Marquard, A.~V.~Smirnov, V.~A.~Smirnov, M.~Steinhauser and D.~Wellmann,
``$\overline{\rm MS}$-on-shell quark mass relation up to four loops in QCD and a general SU$(N)$ gauge group,''
Phys. Rev. D \textbf{94}, 074025 (2016).

\bibitem{Actis:2008ts}
S.~Actis, G.~Passarino, C.~Sturm and S.~Uccirati,
``NNLO Computational Techniques: The Cases $H\rightarrow\gamma\gamma$ and $H\rightarrow gg$,''
Nucl. Phys. B \textbf{811}, 182 (2009).

\bibitem{FCC:2018vvp}
A.~Abada \textit{et al.} [FCC],
``FCC-hh: The Hadron Collider: Future Circular Collider Conceptual Design Report Volume 3,''
Eur. Phys. J. ST \textbf{228}, 755 (2019).

\bibitem{Bi:2015wea}
H.~Y.~Bi, X.~G.~Wu, Y.~Ma, H.~H.~Ma, S.~J.~Brodsky and M.~Mojaza,
``Degeneracy Relations in QCD and the Equivalence of Two Systematic All-Orders Methods for Setting the Renormalization Scale,''
Phys. Lett. B \textbf{748}, 13 (2015).

\bibitem{Baikov:2016tgj}
  P.~A.~Baikov, K.~G.~Chetyrkin and J.~H.~Kuhn,
  ``Five-Loop Running of the QCD coupling constant,''
  Phys.\ Rev.\ Lett.\  {\bf 118}, 082002 (2017).

\bibitem{Wang:2013bla}
S.~Q.~Wang, X.~G.~Wu, X.~C.~Zheng, J.~M.~Shen and Q.~L.~Zhang,
``The Higgs boson inclusive decay channels $H \to b\bar{b}$ and $H \to gg$ up to four-loop level,''
Eur. Phys. J. C \textbf{74},2825 (2014).

\bibitem{Brodsky:2011ta}
 S.~J.~Brodsky and X.~G.~Wu,
 ``Scale Setting Using the Extended Renormalization Group and the Principle of Maximum Conformality: the QCD Coupling Constant at Four Loops,''
 Phys.\ Rev.\ D {\bf 85}, 034038 (2012).

\bibitem{Brodsky:2011ig}
 S.~J.~Brodsky and L.~Di Giustino,
 ``Setting the Renormalization Scale in QCD: The Principle of Maximum Conformality,''
 Phys.\ Rev.\ D {\bf 86}, 085026 (2012).

\bibitem{Brodsky:2012sz}
 S.~J.~Brodsky and X.~G.~Wu,
 ``Application of the Principle of Maximum Conformality to Top-Pair Production,''
 Phys.\ Rev.\ D {\bf 86}, 014021 (2012).

\bibitem{Brodsky:2012rj}
 S.~J.~Brodsky and X.~G.~Wu,
 ``Eliminating the Renormalization Scale Ambiguity for Top-Pair Production Using the Principle of Maximum Conformality,''
 Phys.\ Rev.\ Lett.\ {\bf 109}, 042002 (2012).

\bibitem{Mojaza:2012mf}
 M.~Mojaza, S.~J.~Brodsky and X.~G.~Wu,
 ``Systematic All-Orders Method to Eliminate Renormalization-Scale and Scheme Ambiguities in Perturbative QCD,''
 Phys.\ Rev.\ Lett.\ {\bf 110}, 192001 (2013).

\bibitem{Brodsky:2013vpa}
 S.~J.~Brodsky, M.~Mojaza and X.~G.~Wu,
 ``Systematic Scale-Setting to All Orders: The Principle of Maximum Conformality and Commensurate Scale Relations,''
 Phys.\ Rev.\ D {\bf 89}, 014027 (2014).

\bibitem{Wu:2013ei}
 X.~G.~Wu, S.~J.~Brodsky and M.~Mojaza,
 ``The Renormalization Scale-Setting Problem in QCD,''
 Prog.\ Part.\ Nucl.\ Phys.\ {\bf 72}, 44 (2013).

\bibitem{Wu:2014iba}
 X.~G.~Wu, Y.~Ma, S.~Q.~Wang, H.~B.~Fu, H.~H.~Ma, S.~J.~Brodsky and M.~Mojaza,
 ``Renormalization Group Invariance and Optimal QCD Renormalization Scale-Setting,''
 Rept. Prog. Phys. \textbf{78}, 126201 (2015).

\bibitem{Wu:2019mky}
  X.~G.~Wu, J.~M.~Shen, B.~L.~Du, X.~D.~Huang, S.~Q.~Wang and S.~J.~Brodsky,
  ``The QCD Renormalization Group Equation and the Elimination of Fixed-Order Scheme-and-Scale Ambiguities Using the Principle of Maximum Conformality,''
  Prog.\ Part.\ Nucl.\ Phys.\  {\bf 108}, 103706 (2019).

\bibitem{DiGiustino:2023jiq}
 L.~Di Giustino, S.~J.~Brodsky, P.~G.~Ratcliffe, X.~G.~Wu and S.~Q.~Wang,
 ``High precision tests of QCD without scale or scheme ambiguities,''
 [arXiv:2307.03951 [hep-ph]].

\bibitem{Gell-Mann:1954yli}
M.~Gell-Mann and F.~E.~Low,
``Quantum electrodynamics at small distances,''
Phys. Rev. \textbf{95}, 1300 (1954).

\bibitem{Brodsky:1982gc}
S.~J.~Brodsky, G.~P.~Lepage and P.~B.~Mackenzie,
``On the Elimination of Scale Ambiguities in Perturbative Quantum Chromodynamics,''
Phys. Rev. D \textbf{28}, 228 (1983).

\bibitem{Shen:2017pdu}
J.~M.~Shen, X.~G.~Wu, B.~L.~Du and S.~J.~Brodsky,
``Novel All-Orders Single-Scale Approach to QCD Renormalization Scale-Setting,''
Phys. Rev. D \textbf{95}, 094006 (2017).

\bibitem{Yan:2022foz}
J.~Yan, Z.~F.~Wu, J.~M.~Shen and X.~G.~Wu,
``Precise perturbative predictions from fixed-order calculations,''
J. Phys. G \textbf{50}, 045001 (2023).

\bibitem{Wu:2018cmb}
X.~G.~Wu, J.~M.~Shen, B.~L.~Du and S.~J.~Brodsky,
``Novel demonstration of the renormalization group invariance of the fixed-order predictions using the principle of maximum conformality and the $C$-scheme coupling,''
Phys. Rev. D \textbf{97}, 094030 (2018).

\bibitem{StueckelbergdeBreidenbach:1952pwl}
E.~C.~G.~Stueckelberg de Breidenbach and A.~Petermann,
``Normalization of constants in the quanta theory,''
Helv. Phys. Acta \textbf{26}, 499 (1953).

\bibitem{Peterman:1978tb}
A.~Peterman,
``Renormalization Group and the Deep Structure of the Proton,''
Phys. Rept. \textbf{53}, 157 (1979).

\bibitem{Cacciari:2011ze}
M.~Cacciari and N.~Houdeau,
Meaningful characterization of perturbative theoretical uncertainties,
JHEP \textbf{09}, 039 (2011).

\bibitem{Bagnaschi:2014wea}
E.~Bagnaschi, M.~Cacciari, A.~Guffanti and L.~Jenniches,
An extensive survey of the estimation of uncertainties from missing higher orders in perturbative calculations,
JHEP \textbf{02}, 133 (2015).

\bibitem{Bonvini:2020xeo}
M.~Bonvini,
Probabilistic definition of the perturbative theoretical uncertainty from missing higher orders,
Eur. Phys. J. C \textbf{80}, 989 (2020).

\bibitem{Duhr:2021mfd}
C.~Duhr, A.~Huss, A.~Mazeliauskas and R.~Szafron,
An analysis of Bayesian estimates for missing higher orders in perturbative calculations,
JHEP \textbf{09}, 122 (2021).

\bibitem{Shen:2022nyr}
J.~M.~Shen, Z.~J.~Zhou, S.~Q.~Wang, J.~Yan, Z.~F.~Wu, X.~G.~Wu and S.~J.~Brodsky,
``Extending the Predictive Power of Perturbative QCD Using the Principle of Maximum Conformality and Bayesian Analysis,''
[arXiv:2209.03546 [hep-ph]].

\bibitem{Stevenson:1980du}
  P.~M.~Stevenson,
  ``Resolution of the Renormalization Scheme Ambiguity in Perturbative QCD,''
  Phys.\ Lett.\ B {\bf 100}, 61 (1981).

\bibitem{Stevenson:1981vj}
  P.~M.~Stevenson,
  ``Optimized Perturbation Theory,''
  Phys.\ Rev.\ D {\bf 23}, 2916 (1981).

\bibitem{Ma:2014oba}
  Y.~Ma, X.~G.~Wu, H.~H.~Ma and H.~Y.~Han,
  ``General Properties on Applying the Principle of Minimum Sensitivity to High-order Perturbative QCD Predictions,''
  Phys. Rev. D \textbf{91}, 034006 (2015).

\bibitem{Ma:2017xef}
  Y.~Ma and X.~G.~Wu,
  ``Renormalization scheme dependence of high-order perturbative QCD predictions,''
  Phys. Rev. D \textbf{97}, 036024 (2018).


\bibitem{Callan:1970yg}
C.~G.~Callan, Jr.,
``Broken scale invariance in scalar field theory,''
Phys. Rev. D \textbf{2}, 1541 (1970).

\bibitem{Symanzik:1970rt}
K.~Symanzik,
``Small distance behavior in field theory and power counting,''
Commun. Math. Phys. \textbf{18}, 227 (1970).

\bibitem{Zheng:2013uja}
X.~C.~Zheng, X.~G.~Wu, S.~Q.~Wang, J.~M.~Shen and Q.~L.~Zhang,
``Reanalysis of the BFKL Pomeron at the next-to-leading logarithmic accuracy,''
JHEP \textbf{10}, 117 (2013).

\bibitem{Huang:2021hzr}
X.~D.~Huang, J.~Yan, H.~H.~Ma, L.~Di Giustino, J.~M.~Shen, X.~G.~Wu and S.~J.~Brodsky,
``Detailed comparison of renormalization scale-setting procedures based on the principle of maximum conformality,''
Nucl. Phys. B \textbf{989}, 116150 (2023).

\bibitem{Basdevant:1972fe}
J.~L.~Basdevant,
``The Pade approximation and its physical applications,''
Fortsch. Phys. \textbf{20}, 283 (1972).

\bibitem{Samuel:1992qg}
M.~A.~Samuel, G.~Li and E.~Steinfelds,
``Estimating perturbative coefficients in quantum field theory using Pade approximants. 2.,''
Phys. Lett. B \textbf{323}, 188 (1994).

\bibitem{Samuel:1995jc}
M.~A.~Samuel, J.~R.~Ellis and M.~Karliner,
``Comparison of the Pade approximation method to perturbative QCD calculations,''
Phys. Rev. Lett. \textbf{74}, 4380 (1995).

\bibitem{Du:2018dma}
  B.~L.~Du, X.~G.~Wu, J.~M.~Shen, and S.~J.~Brodsky,
  Extending the Predictive Power of Perturbative QCD,
  Eur.\ Phys.\ J.\ C {\bf 79}, 182 (2019).

\bibitem{TheATLAScollaboration:2015poe}
``Measurement of the Higgs boson production cross section at 7, 8 and 13 TeV center-of-mass energies in the $H\rightarrow\gamma\gamma$ channel with the ATLAS detector,''
ATLAS-CONF-2015-060.

\bibitem{LHCHiggsCrossSectionWorkingGroup:2016ypw}
D.~de Florian \textit{et al.} [LHC Higgs Cross Section Working Group],
``Handbook of LHC Higgs Cross Sections: 4. Deciphering the Nature of the Higgs Sector,''
[arXiv:1610.07922 [hep-ph]].

\bibitem{LHCHiggsCrossSectionWorkingGroup:2013rie}
S.~Heinemeyer \textit{et al.} [LHC Higgs Cross Section Working Group],
``Handbook of LHC Higgs Cross Sections: 3. Higgs Properties,''
[arXiv:1307.1347 [hep-ph]].

\bibitem{ATLAS:2017myr}
[ATLAS],
``Measurements of Higgs boson properties in the diphoton decay channel with 36.1 fb$^{-1}$ $pp$ collision data at the center-of-mass energy of 13 TeV with the ATLAS detector,''
ATLAS-CONF-2017-045.

\bibitem{ATLAS:2022jsi}
[ATLAS],
``Combination of searches for heavy resonances using 139 fb$^{-1}$ of proton\textendash{}proton collision data at $\sqrt{s}$ = 13 TeV with the ATLAS detector,''
ATLAS-CONF-2022-028.

\bibitem{Alves:2022cuz}
F.~L.~Alves [ATLAS],
``Fiducial and differential cross-section measurements in the di-photon channel using full Run2 dataset at ATLAS,''
PoS \textbf{ICHEP2022}, 1051 (2022)

\bibitem{CMS:2017nyv}
[CMS],
``Measurement of differential fiducial cross sections for Higgs boson production in the diphoton decay channel in pp collisions at $\sqrt{s}=13~\mathrm{TeV}$,''
CMS-PAS-HIG-17-015.
	
\bibitem{CMS:2015qgt}
V.~Khachatryan \textit{et al.} [CMS],
``Measurement of differential cross sections for Higgs boson production in the diphoton decay channel in pp collisions at $\sqrt{s}=8\,\text {TeV} $,''
Eur. Phys. J. C \textbf{76}, 13 (2016).

\bibitem{CMS:2022wpo}
[CMS],
``Measurement of the Higgs boson inclusive and differential fiducial production cross sections in the diphoton decay channel with pp collisions at $\sqrt{s}$ = 13 TeV,''
[arXiv:2208.12279 [hep-ex]].

\bibitem{Wang:2016wgw}
S.~Q.~Wang, X.~G.~Wu, S.~J.~Brodsky and M.~Mojaza,
``Application of the Principle of Maximum Conformality to the Hadroproduction of the Higgs Boson at the LHC,''
Phys. Rev. D \textbf{94}, 053003 (2016).


\end{thebibliography}
\end{document}